\documentstyle[prl,aps,epsfig,twocolumn]{revtex}

\begin{document}
\preprint{}
\draft
\title{ Dynamics of Quantum Phase Transition
        in an Array of Josephson Junctions }
\date{October 30, 2001}
\author{ J. Dziarmaga$^{1,2}$,
         A. Smerzi$^{1,3}$,
         W.H. Zurek$^{1}$,
         and A.R. Bishop$^{1}$
}
\address{
 1) Los Alamos National Laboratory, Theory Division,
    Los Alamos, NM 87545, USA\\
 2) Instytut Fizyki Uniwersytetu Jagiello\'nskiego,
    Reymonta 4, 30-059 Krak\'ow, Poland\\
 3) Istituto Nazionale di Fisica per la Materia and
    International School for Advanced Studies,\\
    via Beirut 2/4, I-34014, Trieste, Italy \\
}

\maketitle

\begin{abstract}

We study the dynamics of the Mott insulator-superfluid quantum phase
transition in a periodic 1D array of Josephson junctions. We show that
crossing the critical point at a finite rate with a
quench time $\tau_Q$ induces finite quantum fluctuations of the current
around the loop proportional to $\tau_Q^{-1/6}$. This scaling could be
experimentally verified with in array of weakly coupled Bose-Einstein
condensates or superconducting grains.

\end{abstract}
\pacs{74.50.+r,11.30.Qc,03.75.Fi}

 Thermal fluctuations critically slow down close to a continuous
symmetry breaking thermodynamic phase transition. Both the relaxation
time $\tau$ and the correlation length $\xi$ diverge when a
dimensionless parameter $\epsilon$, which measures the distance from
the critical point, tends to zero: $\xi\sim|\epsilon|^{-\nu}$ and
$\tau\sim|\epsilon|^{-z\nu}$, with $z$ and $\nu$ critical
exponents. In the infinitesimally slow transition the symmetry broken
phase is entered in a state of equilibrium with a fully ordered
complex order parameter. However, in a quench occuring at a finite rate
transition the critical slowing down ($\tau\to\infty$ when
$\epsilon\to 0$) implies that the system goes out of equilibrium some
time before the transition. As a result the complex order parameter
which emerges after the transition will assume different random
phases in different domains of space. This leads to formation of
topological defects such as vortices wherever the circulation of the
phase around a closed loop happens to be nonzero. This process is
known as Kibble-Zurek mechanism (KZM) \cite{kibble76,zurek85,zurek96}. 
The size of the domains follows from a simple argument \cite{zurek85}: 
Close to the transition one can linearize the
time dependence of $\epsilon$ as $\epsilon(t)\approx t/\tau_Q$. The
transition rate is $r(t)=\dot{\epsilon}/\epsilon=1/|t|$. The system
goes out of equilibrium at the time $-\hat{t}$ before the transition
when $r(-\hat{t})=\tau^{-1}(-\hat{t})$. After that time the state of the
system essentially does not change until $+\hat{t}$, when the rate
$r(t)$ becomes again equal to the relaxation rate,
$r(+\hat{t})=\tau^{-1}(+\hat{t})$. At
$\hat{t}=\tau_Q^{z\nu/(1+z\nu)}$ fluctuations of the order parameter
with wavelengths longer than
$\hat{\xi}=\xi(\hat{t})=\tau_Q^{\nu/(1+z\nu)}$ begin to grow
exponentially while short wavelength fluctuations remain unchanged.
The size of the correlated domains is given by $\hat{\xi}$ and the
density of vortices by $1/\hat{\xi}^2$.

  Theoretical and experimental studies of KZM have concentrated so far
on thermal continuous phase transitions with the dynamics of the
order parameter governed by an effective irreversible
time-dependent Ginzurg-Landau theory. Most attention has been
devoted to the normal-superfluid transition in $^3He$
\cite{ruutu96}, superconductors
\cite{kavoussanaki00}, and, more recently, dilute Bose-Einstein
condensates \cite{anglin01}. To date, truly microscopic quantum
approaches have been too complicated to extract useful
predictions.

  A quench-induced {\it quantum} phase transition (QPT) at temperature
$T=0$ must be treated in a microscopic way. It is a common wisdom that some
properties of a quantum transition can be obtained by an exact map
from a thermodynamic transition \cite{sachdev99}: the correlation
length in the ground state of the quantum system scales like
$\xi\sim|\epsilon|^{-\nu}$ and the gap between the ground state and
the first excited state like $dE\sim|\epsilon|^{-z\nu}$. To study the
dynamics of a quantum transition, which drives the system out of its
ground state, we also need information about correlations in excited
states, which is not provided by this map. As we will see below, the
essence of the KZM, which is the competition between the transition
rate and the timescale on which the system can react, is applicable
to quantum transitions. However, the quantum scenario and in
particular interpretation of its results differ from the
thermal case. The main reason is the reversibility of the quantum
dynamics, as opposed to the dynamical irreversibility of the thermal
critical dynamics.

  In this Letter, we study the appearance of a nonzero current while
the system is undergoing a (zero-temperature) quantum phase
transition. We develop a microscopic, dynamical theory and suggest
experiments to test our predictions.

{\bf Josephson junction arrays.} A prototype system displaying a
continuous QPT is an array of mesoscopic Josephson junctions (JJ).
Superconducting JJ networks are now available thanks to important
advances in techniques for ultra-small superconducting
grains although existence of gauge fields may complicate the analysis
\cite{zurek96}. On the other hand, manipulation of optically trapped
Bose-Einstein condensates promises the observation of QPT in neutral
quantum fluids \cite{anderson98,orzel01}. In such systems, a
superconducting/superfluid-insulator phase transition is driven by
the competition between two physical magnitudes: the Josephson
coupling energy $E_J$, which governs the tunneling through the
intra-well barriers, and the on-site interparticle interaction energy
$E_C$. When they are comparable, there is a competition between
long-range order (which is favoured by $E_J$) and localization
(induced by $E_C$). When the latter prevails, no net current can flow
through the junctions.

  An effective Hamiltonian describing the quantum dynamics of an
array of JJ's is given by the Quantum Phase Model (QPM)
\cite{simanek94}:

\begin{equation}
\label{eq1}
\hat{H} = \sum_{l=1}^{N_s} \frac{E_C}{2} \hat{n}_l^2 -
\sum_{\langle k,l \rangle} E_J \cos(\hat{\phi}_k - \hat{\phi}_l)
\end{equation}
with the last sum running over the nearest-neighbour sites. For our
purposes, it is enough to consider the case of a diagonal charging
energy matrix $E_c$, independent of the site indices. The phase
$\hat{\phi}_l$ and the number of atoms $\hat{n}_l$ in each site of
the junction are (with some important caveats \cite{anglin})
non-commuting conjugate observables $[\hat{n}_l,\hat{\phi}_l] = i$,
and, in the $\phi$-representation, $\hat{n}_l \equiv
n-i{\partial
\over {\partial \phi_l}};~ \hat{\phi}_l \equiv \phi_l$. Therefore,
$\phi_l$ and $n_l$ play the role of coordinate and conjugate
momentum, and satisfy the Heisenberg uncertainty relation
The insulator phase is
characterized by large quantum phase fluctuations in the ground
state, which destroy the long-range order among sites. The 1D
Hamiltonian Eq.(\ref{eq1}) exhibits a continuous Mott phase
transition at the critical value $G_c\equiv E_J/E_C=0.617$
\cite{sachdev99}. In what follows we study the microscopic
dynamics of quantum phase transitions in a
one-dimensional chain with periodic boundary conditions.

  The dynamics is governed by the QPM Hamiltonian Eq.(\ref{eq1}), and
satisfies a Schr\"odinger equation

\begin{equation}
i\frac{\partial}{\partial t}\Psi=
-\frac12\sum_{l=1}^{N_s}
\frac{\partial^2}{\partial\phi_l^2} \Psi
-G\sum_{\langle k,l \rangle} \cos(\phi_k-\phi_l) \Psi
 \;\;.
\label{eq2}
\end{equation}
with $G=E_J/E_c$ and the time rescaled as $t \to E_c t$. This
equation is valid for a large number of atoms per site, $n\gg 1$, and
for $G\ll n^2$ \cite{cos2phi}. The ground state below the transition
($G<G_c$) is
close to the Fock state $|n,n,n,\dots\rangle$ with all sites occupied
by the same number of atoms $n$. In the phase representation this
state is described by a uniform wavefunction, $\Psi(\phi_l)={\rm
const}$, where all phase differences between nearest neighbor sites
have the maximal dispersion of $\Delta\phi\sim 1$. Above the
transition $(G>G_c)$ the ground state is a number squeezed state
which continuously tends to a Fock state when $G\rightarrow G_c^+$,
and to a coherent state for $G\gg n^2$. Therefore, when $G\gg G_c$
one can describe the low energy part of the spectrum of
Eq.(\ref{eq2}) in a harmonic approximation \cite{juha}

\begin{equation}
i\frac{\partial}{\partial t}\Psi=
-\frac12
\sum_{l=1}^{N_s}
\frac{\partial^2}{\partial\phi_l^2}
\Psi
+\frac{G}{2}
\sum_{\langle k,l \rangle}
(\phi_k-\phi_l)^2
\Psi.
\label{HO}
\end{equation}
This equation is diagonalized by normal modes numbered by momentum
$\mu\in\{-N_s+1,\dots,+N_s\}$, $\Psi=\prod_{\mu}\Psi_{\mu}(\Phi_{\mu})$.
There is one zero mode $\Phi_0\sim\sum_l\phi_l$. All other modes $\Phi_\mu$
have nonzero frequencies $\sqrt{\gamma_{\mu} G}$[$\approx
G^{1/2}(\frac{2\pi|\mu|}{N_s})$ for small $\mu$], which scale as
$G^{1/2}$,

\begin{equation}
i\frac{\partial}{\partial t}\Psi_{\mu}=
-\frac12
\frac{\partial^2}{\partial\Phi_{\mu}^2}
\Psi_{\mu}+
\frac{\gamma_{\mu}G}{2}
\Phi_{\mu}^2
\Psi_{\mu}.
\label{HOmu}
\end{equation}

  To implement quench with a quench timescale $\tau_Q$ we linearly
ramp the control parameter $G$ in Eq.(\ref{eq2}) as

\begin{equation}
G(t)\;=\;
\frac{t}{\tau_Q} \;\;.
\label{ramp}
\end{equation}
The time $t$ runs from $0$ to $\tau_Q G_{\rm max}$ with $G_{\rm max}\gg G_c$.

{\bf Adiabatic transition.} In the limit $\tau_Q\rightarrow\infty$
the transition is adiabatic and the system state adiabatically
follows its ground state from the Fock state $|n,\dots,n\rangle$ at
$G=0$ to a coherent state with all the nearest neighbor phase
differences close to zero, $\Delta\phi \approx 0$, for $G_{\rm
max}\gg G_c$. The ground state at $G_{\rm max}\gg G_c$ is well
described by the ground state of the harmonic oscillators
(\ref{HOmu}). The dispersion of phases in any harmonic oscillator
ground state at $G_{\rm max}$ is proportional to $G_{\rm
max}^{-1/4}$. The final dispersion of phase differences in the
adiabatic transition is

\begin{equation}
\Delta \phi_{\tau_Q\to\infty} \;\simeq\;G_{\rm max}^{-1/4} \;.
\label{DLinfty}
\end{equation}

{\bf Instantaneous transition.} In the opposite limit of an istantaneous
transition, when $\tau_Q\rightarrow 0$, the system is not able to adjust
its quantum state to the changing hopping rate $G(t)$, and it remains
in the initial Fock state till $G_{\rm max}$. The final dispersion
of the phase differences in the uniform Fock state, $\Psi(\phi_l)={\rm
const}$, is

\begin{equation}
\Delta \phi_{\tau_Q\to 0} \;\sim\; 1 \;.
\label{Fock}
\end{equation}

  For a generic quench the evolution of the system is approximately
adiabatic when $G$ is far from $G_c$, and impulse when $G$ is close to
$G_c$; compare Fig.1 for a 3-site periodic lattice. This is similar to the
thermal case \cite{zurek85}. As $G$ increases starting from $0$ the system 
initially follows its ground state, which is an incoherent
superposition of Fock states, until a $\hat{G}_-< G_c$ when the
transition ceases to be adiabatic. Between $\hat{G}_-$ and a certain
$\hat{G}>G_c$ the evolution is impulse: to a first approximation the state
of the system does not change. The system arrives at $\hat{G}$ in an
incoherent superposition of Fock states with a dispersion
$\Delta\phi\sim 1$ as in the initial Fock state. After $\hat{G}$ the
transition again becomes adiabatic.

{\bf Fast transitions with $(E_c)\tau_Q\ll 1$.} Let us estimate
$\hat{G}$ for a dimensionless $\tau_Q\ll 1$ when we expect that
$\hat{G}\gg G_c$ and we can use the harmonic approximation
(\ref{HO},\ref{HOmu}) for $G\approx\hat{G}$. For the linear quench
(\ref{ramp}) the periods of the oscillators decay like
$G(t)^{-1/2}=\sqrt{\tau_Q/t G_{\rm max}}$. The transition timescale
is $G/\dot{G}=t$. This rate becomes comparable to the periods of the
oscillators at the time $\hat{t}$ when $G^{-1/2}\simeq G/\dot{G}$:
compare Fig.1 for a 3-site periodic lattice. A solution of this
equation gives

\begin{eqnarray}
&&
\hat{t}_{\tau_Q\ll 1} \;\simeq\; \tau_Q^{1/3}
(\;\;{\rm  or  }\;\; E_c^{-2/3}\tau_Q^{1/3})\;,
\label{hatt}\\
&&
\hat{G}_{\tau_Q\ll 1} \;\equiv\;
G(\hat{t}_{\tau_Q\ll 1}) \;\simeq\;
\tau_Q^{-2/3}\;.
\label{hatGamma}
\end{eqnarray}
As anticipated, when $\tau_Q\ll 1$ we have $\hat{G}_{\tau_Q\ll
1}\gg G_c$.

\begin{figure}[h]
\vspace*{-3.0truecm}
\epsfig{width=.9 \linewidth,file=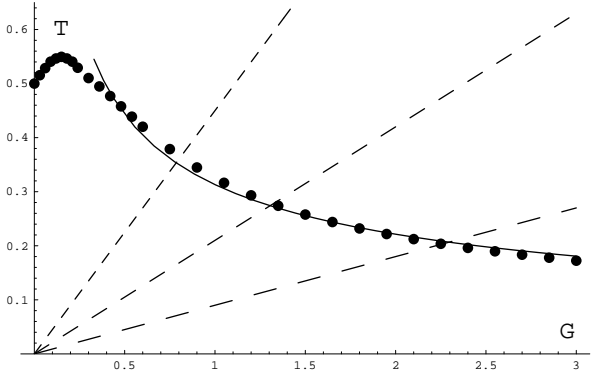}
\vspace{-3.2truecm}
\vskip .3cm
\caption{ The inverse of the gap $T=1/dE$ as a function of $G$ for a
3-site periodic Hubbard model with $n=6$ atoms per site (dots). The solid
line is the best $G^{-1/2}$ fit. The dashed lines show the transition time
$G/\dot{G}=t=G\tau_Q/G_{\rm max}$ for $G_{\rm max}=10$ and
$\tau_Q=4.5,2.1,0.9$ (from left to right). The crossings between $1/dE$
and the dashed lines define $\hat{G}$ for the different $\tau_Q$.}
\label{figure1}
\end{figure}

  We can use the fact that $\hat{G}_{\tau_Q\ll 1}$ to estimate the final
dispersion $\Delta\phi$ at $G_{\rm max}$. At
$\hat{G}_{\tau_Q\ll 1}$ the dispersion is $\Delta\phi\sim 1$. For
$G>\hat{G}_{\tau_Q\ll 1}$ the evolution of the harmonic
oscillators is adiabatic. There is no mixing between the eigenstates of
any oscillator $\mu$ (\ref{HOmu}). The dispersion of the phase
$\Phi_{\mu}$ in any given eigenstate of the oscillator (\ref{HOmu})
scales as $G^{-1/4}$. In the adiabatic evolution for
$G(t)>\hat{G}_{\tau_Q\ll 1}$ the dispersion of $\Phi_{\mu}$
shrinks like $[\hat{G}_{\tau_Q\ll 1}/G(t)]^{1/4}$. $\Delta\phi$
shrinks in the same way. The final dispersion at $G_{\rm max}$ is
$\Delta\phi\simeq[\hat{G}_{\tau_Q\ll 1}/G_{\rm max}]^{1/4}$ or

\begin{equation}
\Delta\phi_{\tau_Q\ll 1}\;\simeq\;
\tau_Q^{-1/6}
G_{\rm max}^{-1/4}
(\;\;{\rm  or  }\;\;
E_c^{-1/6}\tau_Q^{-1/6}G_{\rm max}^{-1/4})\;.
\label{Eq10}
\end{equation}
In a 1D periodic lattice this $\Delta\phi_{\tau_Q\ll 1}$
translates into a dispersion of the angular momentum
$L=i\sum_l(a^{\dagger}_la_{l+1}-a^{\dagger}_{l+1}a_l)$ through the
formula

\begin{equation}
\Delta L \;\sim\; \sqrt{N_s} \; n \; \Delta\phi
\end{equation}
This $\Delta L$ may translate into dispersion of the winding number
$\sqrt{N_s}\Delta\phi$ when the atoms after the quench are forced to
condense. Eq.(\ref{Eq10}) is valid when the predicted $\hat{G}_{\tau_Q\ll
1}\simeq\tau_Q^{-2/3}$ is much less than the final $G_{\rm max}$. This
condition is satisfied when $\tau_Q\gg G_{\rm max}^{-2/3}$. In fact for the
slowest $\mu=\pm 1$ modes to become adiabatic before the maximal possible
$G_{\rm max}=n^2$ we need $\tau_Q\gg n/N_s$.

  Figures 2 and 3 show results of numerical simulations of the 3-site
periodic lattice with a total of $300$ atoms. In Fig.2 we show that
$\Delta\phi$ remains the same as in the initial Fock state for
$G<\hat{G}_{\tau_Q\ll 1}$. In Figure 3 we verify the scaling
$\hat{t}\sim\tau_Q^{1/3}$. We also performed similar simulations for a total
of $120$ (commensurate) and $120+1$ atoms (non-commensurate density of
atoms). Plots like those in Fig.2,3 for $120$ and $121$ atoms are impossible
to distinguish which shows that an extra noncommensurate atom does not make
any difference for fast transitions.

\begin{figure}[h]
\vspace*{-3.0truecm}
\epsfig{width=.9 \linewidth,file=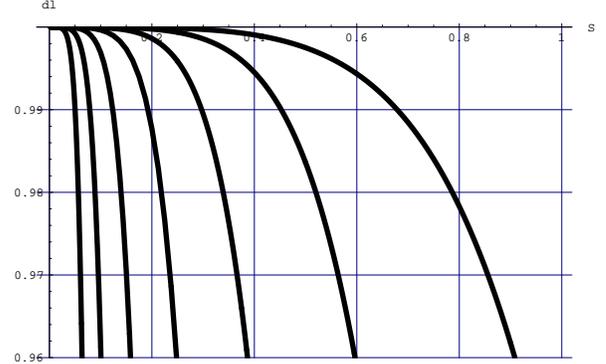}
\vspace{-3.2truecm}
\vskip .3cm
\caption{ The normalized dispersion $dl=\Delta L/\Delta
L_{\tau_Q\to 0}$ as a function of $S=G(t)/G_{\rm max}$. We show
results of exact numerical simulations of the 3-site Hubbard model
for 7 different
$\tau_Q=0.016,0.008,0.004,0.002,0.001,0.0005,0.00025$ (from left
to right) with $G_{\rm max}=1000$ and $n=100$ atoms per site. $dl$
deviates from $1$ (and hits the bottom of the figure) at
$\hat{G}\sim\tau_Q^{-2/3}$: compare Fig.2. }
\label{figure2}
\end{figure}

\begin{figure}[h]
\vspace*{-3.0truecm}
\epsfig{width=.9 \linewidth,file=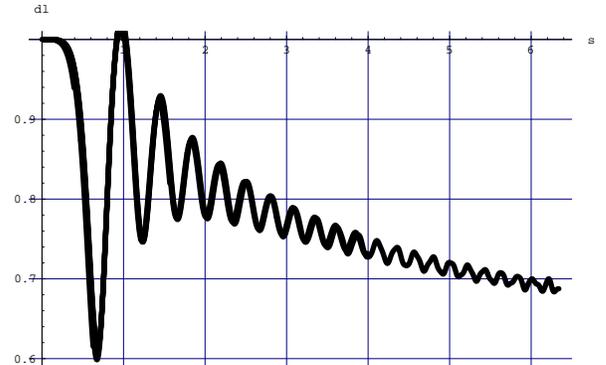}
\vspace{-3.2truecm}
\vskip .3cm
\caption{ The normalized dispersion $dl=\Delta L/\Delta L_{\tau_Q\to 0}$
as a function of $s=G(t)/\hat{G}$ with $\hat{G}\sim\tau_Q^{-2/3}$. This
figure displays {\it all} the seven plots from Fig.2. The plots in Fig.2
are in the top left corner of this figure. The fact that the seven plots
sit on top of each other proves that the evolution of $\Delta L$ depends
on time $t$ through the combination $t/\hat{t}$ with
$\hat{t}\sim\tau_Q^{1/3}$. }
\label{figure3}
\end{figure}

  {\bf Slow transitions with $(E_c)\tau_Q\gg 1$.} So far we concentrated
on $\tau_Q\ll 1$ with $\hat{G}$ well in the harmonic regime. In slower
transitions $\hat{G}$ is close to $G_c$ where the harmonic approximation
cannot be applied. The slow transitions ($\tau_Q\ll 1$) are probing the
critical behaviour close to $G_c$.

  In the perioric $3$-site array there is no phase transition but,
rather, a crossover: the gap $dE$ is minimal at the crossover point
$G_c\approx 0.5$ but it does not vanish there, compare Fig.1. For
$\tau_Q\gg 1$ the transition is adiabatic and the system follows its
ground state. The phase difference dispersion at $G_{\rm
max}\gg G_c$ is the dispersion in the ground state of the harmonic
oscillators (\ref{DLinfty}). A thermal crossover transition was studied
in \cite{APP}.

  In a large $2\;$D array with a commensurate density of atoms
there is a phase transition and not a crossover
\cite{sachdev99,fisher}. The gap $dE$ vanishes at $G=G_c$.  Close
to the critical point, $G\approx G_c$, the energy gap scales as
$dE\sim (G-G_c)^{\nu}$, where $\nu=2/3$ is an exact
renormalization group critical exponent. \thanks{ Note that in the
harmonic regime for $G\gg G_c$ we have $dE\sim G^{1/2}$ which is
consistent with a mean-field value of $\nu=1/2$.} With a
linearized $G(t)-G_c=\frac{t}{\tau_Q}$ the gap $\Delta E\sim
(G-G_c)^{2/3}=(t/\tau_Q)^{2/3}$ becomes equal to the quench rate
$r(t)=\dot{G}/(G-G_c)=1/t$ at $\hat{G}\simeq G_c+1/\tau_Q^{3/5}$.
Before $\hat{G}$ the system is in an incoherent superposition of
Fock states. After $\hat{G}$ the evolution is adiabatic, and with
increasing $G$ the phase dispersion shrinks together with the
phase widths of the system eigenstates. The dispersion measured at
$G_{\rm max}\gg G_c$, where we can use the harmonic approximation,
is

\begin{equation}
\Delta \phi_{\tau_Q\gg 1}^{\rm 2D}\;\approx\;
\frac{    \hat{G}^{1/4}    }
     {  G_{\rm max}^{1/4}  } \;\simeq\;
\frac{  (G_c+\frac{1}{(E_c^{3/5})\tau_Q^{3/5}})^{1/4}  }
     {   G_{\rm max}^{1/4}                  }\;.
\end{equation}
This formula is consistent with Eq.(\ref{DLinfty}) because $G_c={\cal
O}(1)$. Here we again use the fact that the phase dispersion of harmonic
oscillator eigenstates shrinks like $G^{-1/4}$. The critical behaviour
is realized in the $(E_c^{3/5})\tau_Q^{3/5}$ term.

{\bf Concluding remarks.} The state of the system after the transition
does not have definite angular momentum or definite phase differences
between lattice sites. That would be the case in the thermal KZM, where
$\Delta\phi$ would describe the dispersion in an ensemble of different
possible classical outcomes. In a quantum phase transition at zero
temperature the state of the system after the transition is in a {\it
coherent} superposition of states with different $\Delta\phi$.  Either a
measurement or decoherence \cite{zurek91} are needed to convert this coherent 
superposition into a mixtures of states, each with definite current.

  Another difference with respect to the thermal KZM is due to the
reversibility of the quantum dynamics. In the thermal case the characteristic
lenghtscale $\hat{\xi}$ is frozen after the symmetry breaking transition
is completed. $\hat{\xi}$ is a permanent record of the transition rate
$\tau_Q$: manipulations with $\epsilon$ do not change the winding number
as long as the system remains in the symmetry broken phase with
$\epsilon>0$. In contrast, even after the diabatic quantum transition is
completed one can change $\Delta\phi_{\tau_Q}$. Adiabatic variations of
$G$ away from $G_{\rm max}$ are accompanied by changes in $\Delta\phi$,
$\Delta\phi=\Delta\phi_{\tau_Q}(G_{\rm max}/G)^{1/4}$.

  In conclusion, we have predicted the phase dispersion after a diabatic
insulator-superfluid quantum phase transition in an array of Josephson
junctions. This theory is a quantum counterpart of the Kibble-Zurek
mechanism for topological defect formation in classical thermal phase
transitions. Our predictions can be tested experimentally in 1D or 2D
superconducting JJ arrays. The possibility of tuning the Josephson
coupling energy in a superconducting JJ has been demonstrated recently
\cite{schon01}. 1D (or 3D \cite{3D}) JJ arrays realized with Bose-Einstein
condensates can be, perhaps even more easily, accurately tailored
\cite{anderson98,orzel01,cataliotti01,trombettoni01}. For the case of a 1D
array our Eq.(\ref{Eq10}) predicts a dispersion of the phases
$\Delta\phi\simeq G_{\rm max}^{-1/4} \tau_Q^{-1/6}$ that can be directly
measured by the interference techniques which have been used in the
experiments \cite{anderson98,orzel01,cataliotti01}.

We acknowledge useful discussions with James
Anglin and Diego Dalvit. This work was partially supported by the DOE
and the Cofinanziamento MURST.


\begin{references}

\bibitem{sachdev99} S.~Sachdev, Quantum Phase
Transitions, Cambridge University Press, (1999).

\bibitem{kibble76} T.W.B.~Kibble, J. Phys. {\bf A9}, 1387
(1976); T.W.B.~Kible and A.~Vilenkin, Phys. Rev. {\bf
D52}, 679 (1995).

\bibitem{zurek85} W.H.~Zurek, Nature (London)  {\bf 317}, 505 (1985).

\bibitem{zurek96} W.H.~Zurek, Phys. Rep. {\bf 276}, 177 (1996);
M.~Hindmarsh and A.~Rajantie, Phys.Rev.Lett.{\bf 85}, 4660 (2000);
G.J.~Stephens, L.M.A.~Bettencourt, and W.H.~Zurek, cond-mat/0108127.

\bibitem{kavoussanaki00} R.~Carmi {\it et al.}, Phys. Rev. Lett.{\bf 84},
4966 (2000); E.~Kavoussanaki {\it et al.}, Phys. Rev. Lett.{\bf 85}, 3452
(2000).

\bibitem{ruutu96} V.M.H.~Ruutu et al, Nature {\bf 382}, 334
                  (1996); Bauerle {\it et.al.},
                  Nature {\bf 382}, 332 (1996).

\bibitem{anglin01} J.R.~Anglin and W.H.~Zurek,
                   Phys.Rev.Lett. {\bf 83}, 1707 (1999).

\bibitem{sondhi97} S.L.~Sondhi, S.M.~Girvin, J.P.~Carini, D.~Shahar,
Rev of Mod. Phys. {\bf 69}, 315 (1997)

\bibitem{barone82} A.~Barone and G.~Paterno, {\it Physics and
Applications of the Josephson Effect}, Wiley, New York, 1982.

\bibitem{anderson98} B.P.~Anderson and M.A.~Kasevich,
Science {\bf 282}, 1686 (1998).

\bibitem{orzel01} C.~Orzel, A.K.~Tuchman, M.L.~Fenslau,
M.~Yasuda and M.A.~Kasevich, Science {\bf 291}, 2386 (2001).

\bibitem{simanek94} E.~Simanek, {\it Inhomogeneous Superconductors},
Oxford University Press, 1994.

\bibitem{anglin} J.~Anglin, P.~Drummond and A.~Smerzi,
Phys.Rev {\bf A64}, 063605 (2001)

\bibitem{cos2phi} With this assumption, satisfied in the experiment
\cite{orzel01}, we could neglect in Eq.(2) 
the $O(G/n^2)$ correction terms derived in \cite{anglin}.

\bibitem{juha} J.~Javanainen, Phys. Rev. {\bf A60}, 4902
(1999).

\bibitem{schon01} J.H.~Schon, C.~Kloc, H.Y.~Hwang and B.~Batlogg,
Science {\bf 292}, 252 (2001).

\bibitem{cataliotti01} F.S.~Cataliotti {\it et.al.}, Science {\bf 293},
843 (2001).

\bibitem{trombettoni01} A.~Trombettoni and A.~Smerzi,
Phys. Rev. Lett. {\bf 86} 2353 (2001)

\bibitem{APP} W.H.~Zurek, L.M.A.~Bettencourt,
              J.~Dziarmaga, and N.D.~Antunes,
              Acta.Phys.Polon.{\bf B} 32, 2279 (2001);
              see also J.~Dziarmaga, Phys.Rev.Lett.{\bf 81},
              5485 (1998).

\bibitem{fisher} M.P.A.~Fisher, P.B.~Weichman, G.~Grinstein, and
                 D.S.~Fisher, Phys.Rev.{\bf B} 40, 546 (1989).

\bibitem{zurek91} W.H.~Zurek, Phys. Today {\bf 44}, 36 (1991); quant-ph/010527.

\bibitem{3D} M.~Greiner {\it et al.}, Nature {\bf 415}, 39 (2002).

\end{references}
\end{document}